\documentclass[twocolumn,amsmath,amssymb,pra,floatfix]{revtex4}
\usepackage{graphicx}
\usepackage{bm}
\usepackage{amsmath,amsfonts}

\begin{document}

\title{Interacting Fibonacci anyons in a Rydberg gas}

\author{Igor Lesanovsky}

\affiliation{Midlands Ultracold Atom Research Centre (MUARC), School of Physics and Astronomy, The University of Nottingham, Nottingham NG7 2RD, United Kingdom}

\author{Hosho Katsura}
\affiliation{Department of Physics, Gakushuin University, Mejiro, Toshima-ku, Tokyo 171-8588, Japan}

\date{\today}

\begin{abstract}
A defining property of particles is their behavior under exchange. In two
dimensions anyons can exist which, opposed to fermions and bosons, gain arbitrary relative phase factors \cite{Arovas85} or even undergo a change of their type. In the latter case one speaks of non-Abelian anyons - a particularly simple and aesthetic example of which are Fibonacci anyons \cite{Trebst08}. They have been studied in the context of fractional quantum Hall physics where they occur as quasiparticles in the $k=3$ Read-Rezayi state \cite{Read99}, which is conjectured to describe a fractional quantum Hall state at filling fraction $\nu=12/5$ \cite{Xia04}. Here we show that the physics of interacting Fibonacci anyons \cite{Feiguin07} can be studied with strongly interacting Rydberg atoms in a lattice, when due to the dipole blockade \cite{Lukin01} the simultaneous laser excitation of adjacent atoms is forbidden. The Hilbert space maps then directly on the fusion space of Fibonacci anyons and a proper tuning of the laser parameters renders the system into an interacting topological liquid of non-Abelian anyons.  We discuss the low-energy properties of this system and show how to experimentally measure anyonic observables.
\end{abstract}
\maketitle
There is great interest in studying many-body systems of interacting anyons as they often exhibit exotic quantum phases \cite{Feiguin07,Trebst08-2,Gils09-2,Grosfeld09}. A further motivation is that the exchange of anyons - the braiding - permits the implementation of robust protocols for quantum information processing  \cite{Kitaev03,Chetan08,Brennen08}. Physically, anyons emerge as quasi particles on the ground state of an interacting many-body system and recently there has been put much effort in implementing and exploring anyonic models with cold atoms \cite{Paredes01,Aguado08,Weimer10}, polar molecules \cite{Micheli06} and trapped ions \cite{Barreiro11}. A model which has been much studied - in particular in the context of quantum information processing - is Kitaev's toric code \cite{Kitaev06}. Excitations on the ground state are Abelian anyons, which merely acquire a phase when braided. A more complex and exotic scenario is encountered in the non-Abelian case, i.e. when anyons can undergo an actual change of their type under braiding.

\begin{figure}
\includegraphics[width=\columnwidth]{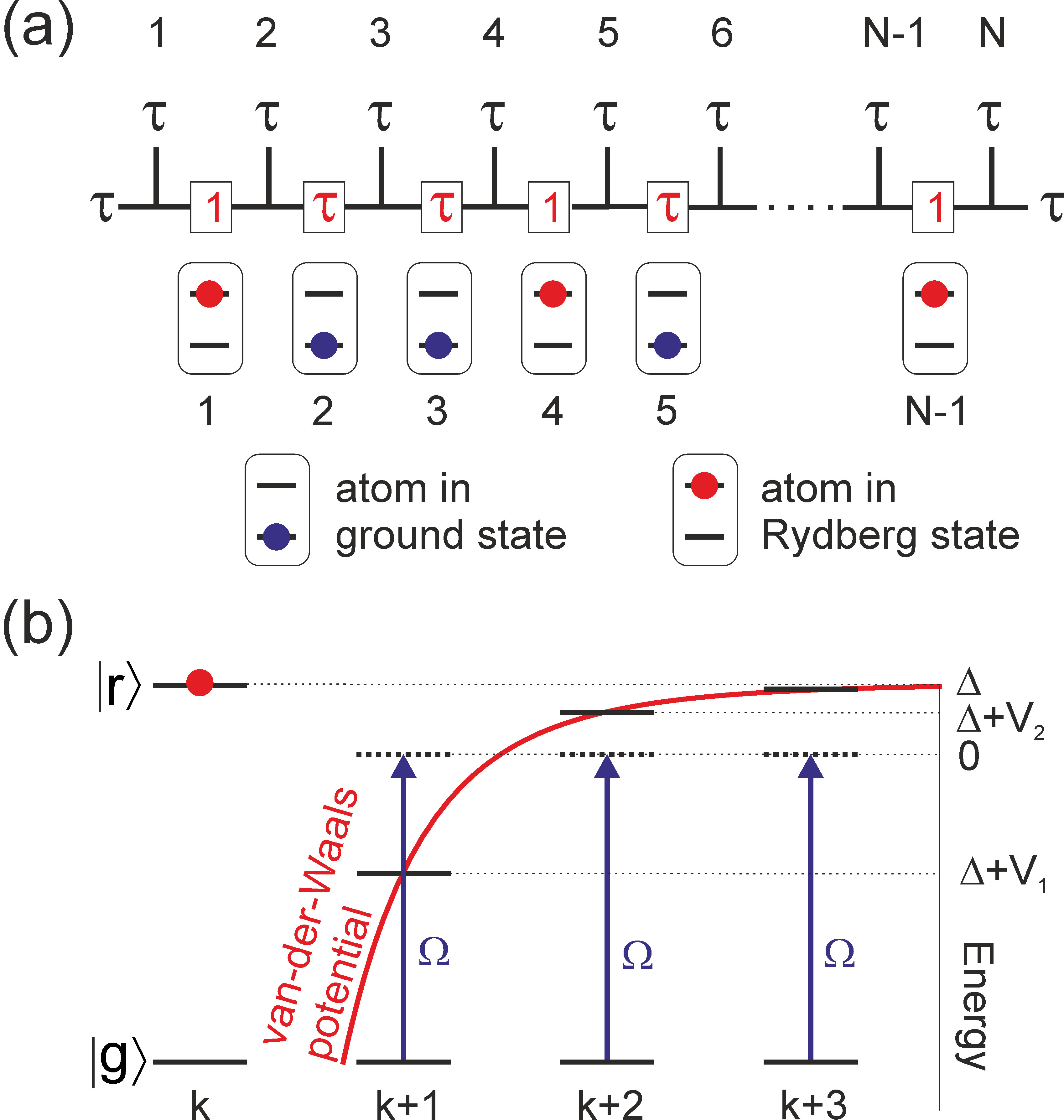}
\caption{\textbf{(a) Fusion space.} The Hilbert space of a system of $N$ Fibonacci anyons of type $\tau$ is spanned by all possible fusion paths. An example of such path is given by the symbols contained in the boxes. A fusion path translates into configurations of atoms which are either in the electronic ground state or excited to a Rydberg state. The Rydberg state (ground state) is identified with the fusion outcome $1$ ($\tau$). \textbf{(b) Level scheme and interaction energy.} The excitation of a Rydberg atom on site $k$ shifts the energy of Rydberg states of atoms in the neighborhood. The strong interaction between neighboring atoms (here $k$ and $k+1$) excludes their simultaneous excitation to Rydberg states and effectively constrains the Hilbert space of the atomic system to the fusion space of Fibonacci $\tau$-anyons.}
\label{fig:fibonacci}
\end{figure}

A particularly simple example of non-Abelian anyons are Fibonacci anyons which occur in two types: A trivial particle referred to as $1$ and a non-trivial particle denoted by $\tau$. A convenient way to define an anyonic system is through \emph{fusion rules} for the particle types (see e.g. Refs. \cite{Brennen08,Trebst08}), which for Fibonacci anyons read
\begin{eqnarray*}
 1 \times 1 = 1,\,\,
 1 \times \tau = \tau,\,\,
 \tau \times 1 = \tau,\,\,
 \tau\times\tau = 1 + \tau.
\end{eqnarray*}
The first three rules are a consequence of the trivial nature of a $1$-anyon. The fourth rule states that two anyons of type $\tau$ can fuse such that the result is either the trivial particle or a $\tau$-anyon. This can be thought of as being analogous to the merging of two spin $1/2$ particles ($\frac{1}{2}\otimes\frac{1}{2}=0\oplus1 $), which can yield either a singlet (total spin $0$) or a triplet (total spin $1$).

For a set of $N$ anyons of type $\tau$ it can be shown that a basis of the underlying Hilbert space is spanned by all possible fusion paths \cite{Trebst08}, i.e. the number of ways in which these anyons can be successively fused. For large $N$, this number grows as $\varphi^N$ with $\varphi=(1+\sqrt{5})/2$ being the \emph{golden ratio}. An example of a fusion path is illustrated in Fig. \ref{fig:fibonacci}a and it is constructed as follows: Starting from the left the first $\tau$-anyon fuses with the $\tau$ that is set by the left boundary condition. There are two possible fusion outcomes, $1$ or $\tau$. In this particular realization we choose the outcome $1$, which is written into the first box. This trivial particle then fuses with the second $\tau$-anyon. The fusion rules dictate that the outcome must be a $\tau$ anyon (second box). Continuing this procedure by obeying the fusion rules results in a possible fusion path. It is important to note that since a trivial particle fusing with a $\tau$ always results in another $\tau$ the occurrence of two consecutive $1$'s is excluded.

In the following we will show in detail that a Rydberg lattice gas constitutes an analog quantum simulator for Fibonacci anyons. The link between these two systems is the aforementioned exclusion principle. While for anyons this is a consequence of the underlying mathematical rules the exclusion in a Rydberg system is physically rooted in the dipole blockade \cite{Lukin01} which prevents the simultaneous excitation of neighboring atoms to Rydberg states. The blockade originates from large electrostatic energy shifts between atoms in Rydberg states and has recently been demonstrated experimentally for atoms trapped in individual optical traps \cite{Urban09,Gaetan09}. Also first experiments with Rydberg atoms in a lattice \cite{Viteau11} have been conducted. We will discuss which anyonic interactions can be realized in such a system and how anyonic degrees of freedom can be measured experimentally. We expect that this comparatively simple quantum simulator platform for non-Abelian anyons will highlight a new route towards the study of exotic forms of quantum matter.

Formally, the Rydberg lattice gas is described by a set of atoms with the electronic ground state $\left|g\right>\equiv\left|\downarrow\right>$ that is coupled by a laser with Rabi frequency $\Omega$ to a high lying Rydberg state denoted as $\left|r\right>\equiv\left|\uparrow\right>$. Excited atoms with position labels $k$ and $m$ interact with a van-der-Waals potential $V_{|k-m|}=(C_6/a^6)/|k-m|^{6}$ where $C_6$ is the interaction dispersion coefficient and $a$ the lattice spacing. Within the rotating-wave approximation the Hamiltonian of an ensemble of $N-1$ atoms reads
\begin{eqnarray}
  H=\Omega \sum_{k=1}^{N-1} \sigma^x_k+\Delta \sum_{k=1}^{N-1} n_k+\sum_{k=1,m\neq k}^{N-1} V_{|k-m|} n_k n_m. \label{eq:hamiltonian}
\end{eqnarray}
Here $\Delta$ is the detuning of the laser frequency with respect to the frequency of the atomic transition $\left|g\right>\leftrightarrow\left|r\right>$, $n_k=(1+\sigma^z_k)/2$ is the local number operator and $\sigma^{x,z}_k$ are Pauli spin matrices referring to the internal degrees of freedom of each atom. Hamiltonian (\ref{eq:hamiltonian}) has been studied in a number of theoretical works \cite{Weimer10-1,Lesanovsky11,Mukherjee11,Lesanovsky12} and recent experiments have shown that it accurately reflects the physics of laser driven Rydberg gases \cite{Low09,Viteau11}.

We are here interested in a parameter regime in which the nearest neighbor interaction is the largest energy scale, i.e. $|\Omega|,|\Delta|\ll |V_1|$. Fig. \ref{fig:fibonacci}b depicts a sketch of this situation where we take $V_1$ ($C_6$) \cite{Mukherjee11} negative. However, in this work we will discuss both positive as well as negative $V_1$. It was shown in Refs. \cite{Lesanovsky11,Lesanovsky12} that when $|\Omega|,|\Delta|\ll |V_1|$ the dynamics of the Rydberg gas is confined to a Hilbert space $\mathcal{H}_\mathrm{blockade}$ that is spanned by atomic configurations in which \emph{adjacent atoms are not simultaneously excited}. This Hilbert space is equivalent to the fusion space of an ensemble of $N$ Fibonacci anyons of type $\tau$. To see this, one should think of the atoms as being located in between the $\tau$-anyons. A fusion path (see Fig. \ref{fig:fibonacci}a for an example) is then encoded in the internal state of the atoms: the $k$-th atom is in the state $\left|r\right>$($\left|g\right>$) if the outcome of the fusion of the $k$-th anyon with the previous fusion outcome is $1$($\tau$). The state of each atom can thus be interpreted as the combined topological charge of the anyons located to its left \cite{Feiguin07}.

Having established a formal equivalence between the Hilbert spaces of a Rydberg lattice gas and the fusion space, we will now turn to the question concerning which anyonic interactions are actually realized by the atomic system. To this end, it is more convenient to work with an explicit projection of Hamiltonian (\ref{eq:hamiltonian}) on the constrained Hilbert space $\mathcal{H}_\mathrm{blockade}$. This effective Hamiltonian reads $H_\mathrm{eff}=H_0+H_\mathrm{vdW}+H_2$ with
\begin{eqnarray*}
  H_0&=&\Omega\left[\sigma_1^x P_2 +P_{N-2}\sigma^x_{N-1} \right]+\Omega \sum_{k=2}^{N-2} P_{k-1}\sigma^x_k P_{k+1}\nonumber\\
  &&+\Delta \sum_{k=1}^{N-1} n_k+V_2\sum_{k=1}^{N-3} n_k n_{k+2}\\
  H_\mathrm{vdW}&=&\sum_{|k-m|>2} \!\!\!\!V_{|k-m|} n_k n_m\\
  H_2&=&-\frac{\Omega^2}{V_1}\left[2\sum^{N-1}_{k=1}  n_k -\frac{3}{2} n_k n_{k+2}\right.\nonumber\\
  &&+\left.\sum^{N-4}_{k=1} P_{k}\left(\sigma^+_{k+1} \sigma^-_{k+2}+\sigma^-_{k+1} \sigma^+_{k+2}\right)P_{k+3}\right],
\end{eqnarray*}
where $P_k=1-n_k$. Here $H_0+H_\mathrm{vdW}$ is the actual projection of Hamiltonian (\ref{eq:hamiltonian}) onto $\mathcal{H}_\mathrm{blockade}$. The term $H_2$ contains second order corrections that arise from a non-perfect blockade, i.e. due to a finite nearest-neighbor interaction. Since the relative strength of $H_2$ scales with $(\Omega/V_1)^2$ it can be regarded as a perturbation to $H_0$. Moreover, due to the quickly decaying tail of the van-der-Waals interaction also the term $H_\mathrm{vdW}$ can be considered perturbatively. We will later discuss the influence of these terms on the spectrum, but let us first analyze the principal Hamiltonian $H_0$. We start by rewriting it as $H_0=\sum_{k=2}^{N-2}h_k+H_\mathrm{b}$ where the $h_k$ are local three body Hamiltonians and $H_\mathrm{b}$ defines the boundary term. Using the basis (centered around the $k$-th site)
$\{\left|\uparrow\downarrow\uparrow\right>, \left|\downarrow\downarrow\uparrow\right>, \left|\uparrow\downarrow\downarrow\right>,\left|\downarrow\uparrow\downarrow\right>,
\left|\downarrow\downarrow\downarrow\right>\}$ and taking into account that these site-triples overlap, we find that
\begin{align}
  h_k=\left(
        \begin{array}{ccccc}
          f_1 &  &  &  &  \\
           & f_2 &  &  &  \\
           &  & \alpha &  &  \\
           &  &  & \beta & \Omega \\
           &  &  & \Omega & \gamma \\
        \end{array}
      \right)_{k}\label{eq:hamiltonian_3sites}
\end{align}
where $f_1=\Delta+V_2-\beta+2\gamma$, $f_2=\Delta-\alpha-\beta+3\gamma$ and $\alpha$, $\beta$ and $\gamma$ are constants. These constants are arbitrary, showing that there is a large class of local three-body Hamiltonians whose sum adds up to Hamiltonian $H_0$. The boundary term reads $H^\mathrm{o}_\mathrm{b}=(\Delta-\alpha+\gamma)n_1+(f_2-\gamma) n_2+(\alpha-\gamma) n_{N-2}+(\Delta-f_2-\gamma)n_{N-1}+\Omega\left[\sigma_1^x P_2 +P_{N-2}\sigma^x_{N-1}\right]$ for open and $H^\mathrm{c}_\mathrm{b}=h_1+h_{N-1}$ for closed boundaries (with relabeling the indices $N\rightarrow 1$, $0\rightarrow N-1$).

An anyonic interaction which has been extensively discussed in the literature \cite{Feiguin07,Trebst08} is the anyonic equivalent of the Heisenberg interaction between spins. As mentioned earlier two spin $1/2$ particles fuse according to the rule $\frac{1}{2}\otimes\frac{1}{2}=0\oplus1$, where $0$ and $1$ are the total spins (singlet and triplet) and the Heisenberg interaction, $\propto \mathbf{S}_1\cdot \mathbf{S}_2$, introduces an energy difference between these fusion outcomes. According to $\tau\times\tau=1 +\tau$ an anyonic analogue thus would assign different energies to the two fusion outcomes $1$ and $\tau$. The construction of this Hamiltonian is discussed in Refs. \cite{Feiguin07,Trebst08}, but we will briefly summarize here how an explicit representation of the interaction in fusion space and hence also in $\mathcal{H}_\mathrm{blockade}$ is obtained.
\begin{figure*}
\includegraphics[width=2\columnwidth]{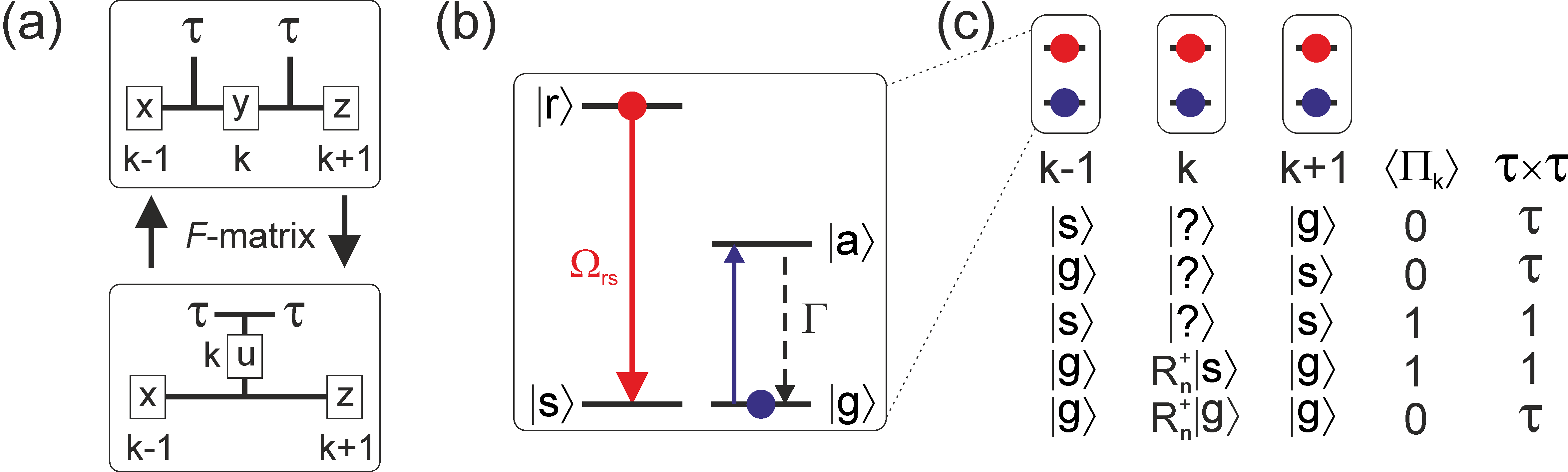}
\caption{\textbf{(a) Changing the order of the fusion path.} The $F$-matrix is performing a basis change in fusion space such that the fusion outcome $u$ of the two neighboring $\tau$-anyons becomes explicit. \textbf{(b) Determination of the fusion outcome via a measurement of the three-point correlation function} $\left<\Pi_k\right>$. We start by applying a strong laser pulse with Rabi frequency $\Omega_\mathrm{rs}\gg |V_2|$ on a transition from the Rydberg state $\left|r\right>$ to a stable hyperfine state $\left|s\right>$. This state transfer is necessary for switching off the Rydberg-Rydberg interaction. We then measure the state of the atoms on sites $k\pm1$ by monitoring fluorescence from a closed transition, i.e. we shine in a laser that resonantly couples the state $\left|g\right>$ to a short lived state $\left|a\right>$ that decays under the emission of a photon (with rate $\Gamma$) back to $\left|g\right>$. If the atom is in state $\left|g\right>$ this results in the cyclic emission of photons while in the opposite case (atom in $\left|s\right>$) no photons are emitted. The presence/absence of these scattered photons thus allows directly to infer the atomic state. \textbf{(c) List of measurement results and the corresponding fusion outcome.} If one finds that the atoms located at sites $k-1$ and $k+1$ are in different internal states we directly conclude from the matrix representation (\ref{eq:projector}) that $\left<\Pi_k\right>=0$ and hence the fusion outcome was $\tau$. In turn, if the atoms are in the state $\left|s\right>_{k-1}\left|s\right>_{k+1}$, we can conclude that $\left<\Pi_k\right>=1$ (fusion of the two $\tau$'s to a trivial anyon). The most involved scenario is encountered when the two outer atoms are in the state $\left|g\right>_{k-1}\left|g\right>_{k+1}$. In this case we have to measure the expectation value of the operator that corresponds to the lower right $2\times2$-block of $\Pi_k$ given in  Eq. (\ref{eq:projector}). This block can be written as a rotation of the number operator $n=(1+\sigma^z)/2$: $R^\dagger_\mathbf{n}\,n\,R_\mathbf{n}$.
Here $R_\mathbf{n}=e^{i\pi \mathbf{n}\cdot\mathbf{\sigma}/2}$ with $\mathbf{\sigma}=(\sigma^x,\sigma^y,\sigma^z)$ conducts a rotation around the axis $\mathbf{n}=(\varphi^{-1/2},0,\varphi^{-1})$ by an angle $\pi$. Hence, if we rotate the state of the $k$-th atom on the Bloch sphere by applying $R_\mathbf{n}$ and subsequently perform a projective measurement we find the fusion outcomes to be $1$($\tau$) if the atom is in the state $\left|s\right>$($\left|g\right>$).}
\label{fig:basis_change_measurement}
\end{figure*}

When considering two $\tau$ anyons there are different orders in which anyons can be fused, two of which are depicted in Fig. \ref{fig:basis_change_measurement}a. The upper panel represents a ``conventional" fusion path according to the rules discussed earlier: Here the left anyon of type $x$ merges with the first $\tau$-anyon. The outcome of this fusion is an anyon of type $y$ which fuses with the second $\tau$ to yield an anyon of type $z$. This is not a convenient basis for the discussion of anyonic interactions, as we cannot directly read off the fusion outcome of the two $\tau$'s. We thus perform a basis change from the ``conventional" fusion space $\left|xyz\right>\,\in\,\{\left|1\tau 1\right>, \left|\tau\tau 1\right>,\left|1\tau \tau\right>,\left|\tau 1 \tau\right>,\left|\tau\tau \tau\right>\}$ to a basis where the fusion outcome (denoted as $u$) of the two $\tau$-anyons becomes explicit: $\left|xuz\right>\,\in\,\{\left|111\right>, \left|\tau\tau 1\right>,\left|1\tau \tau\right>,\left|\tau 1 \tau\right>,\left|\tau\tau \tau\right>\}$. The change to this new basis, as depicted in Fig. \ref{fig:basis_change_measurement}a, is conducted by the so-called $F$-matrix \cite{Feiguin07,Trebst08}
\begin{eqnarray}
  F_k=\left(\begin{array}{ccccc}
  1 & & & & \\
  & 1 & & & \\
  & & 1 & & \\
  & & & \varphi^{-1} & \varphi^{-1/2} \\
  & & & \varphi^{-1/2} & -\varphi^{-1}
  \end{array}\right)_k.
\end{eqnarray}
With this we are now equipped to construct the Heisenberg interaction. An operator that discriminates the two fusion results is a projector on the trivial particle. In the basis where the fusion outcome of the two $\tau$'s is explicit (bottom panel of Fig. \ref{fig:basis_change_measurement}a), this is a diagonal matrix with entries one, where  $u=1$ and zero otherwise, $\bar{\Pi}_k=\mathrm{diag}(1,0,0,1,0)$. Upon transforming $\bar{\Pi}_k$ back into the ``conventional" fusion space we can thus write the anyonic Heisenberg Hamiltonian as $H_\mathrm{aH}=-J \sum_{k=1}^N \Pi_k$ with
\begin{eqnarray}
  \Pi_k=F_k\,\bar{\Pi}_k\,F_k=\left(\begin{array}{ccccc}
  1 & & & & \\
  & 0 & & & \\
  & & 0 & & \\
  & & & \varphi^{-2} & \varphi^{-3/2} \\
  & & & \varphi^{-3/2} & \varphi^{-1}
  \end{array}\right)_k.\label{eq:projector}
\end{eqnarray}
Comparing the coefficients of $\Pi_k$ with those of the three-body Hamiltonians $h_k$ in Eq. (\ref{eq:hamiltonian_3sites}) we find that this interaction is in fact naturally present in a system of interacting Rydberg atoms when
\begin{eqnarray}
  \Omega=-J\,\varphi^{-3/2},\,\Delta=-J\,(\varphi^{-2}-3\varphi^{-1}),\,V_2=-J\,\varphi.
  \label{eq:parameters}
\end{eqnarray}
With this choice of parameters the Rydberg lattice gas behaves just like a system of Fibonacci anyons that interact with a Heisenberg interaction. For $J>0$ ($V_2<0$), it is energetically favorable for anyons to fuse to the trivial particle. In the opposite case, $J<0$ ($V_2>0$), the fusion to a $\tau$-anyon is favored.

Before turning to the analysis of the spectral properties and the influence of imperfections such as a non-perfect Rydberg blockade or the van-der-Waals tail let us show how one can experimentally measure properties of the anyonic system. We have mentioned earlier that the state of an atom encodes the combined topological charge of the anyons located to its left. This property can hence be readily inferred from a site-resolved projective measurement. Another important experimental quantity - in particular in the context of the discussed Heisenberg interaction - is the result of a fusion of two $\tau$-anyons. To determine the fusion outcome of the anyons with label $k$ and $k+1$ one needs to determine the expectation value of the projection operator $\Pi_k$. This amounts to a measurement of a three-point correlation function on the atomic degrees of freedom. If the outcome is $1$($0$) we know that the particles have fused to an anyon of type $1$($\tau$). $\left<\Pi_k\right>$ can be inferred from three separate projective measurements each of which concern only a single site. Such site resolved addressing in conjunction with projective measurements has been experimentally demonstrated in the context of interacting Rydberg atoms \cite{Urban09,Gaetan09}. A schematics of the procedure and involved levels is given in Fig. \ref{fig:basis_change_measurement}b, and a comprehensive list of all measurement outcomes together with the corresponding fusion result is provided in Fig. \ref{fig:basis_change_measurement}c.

\begin{figure}
\includegraphics[width=\columnwidth]{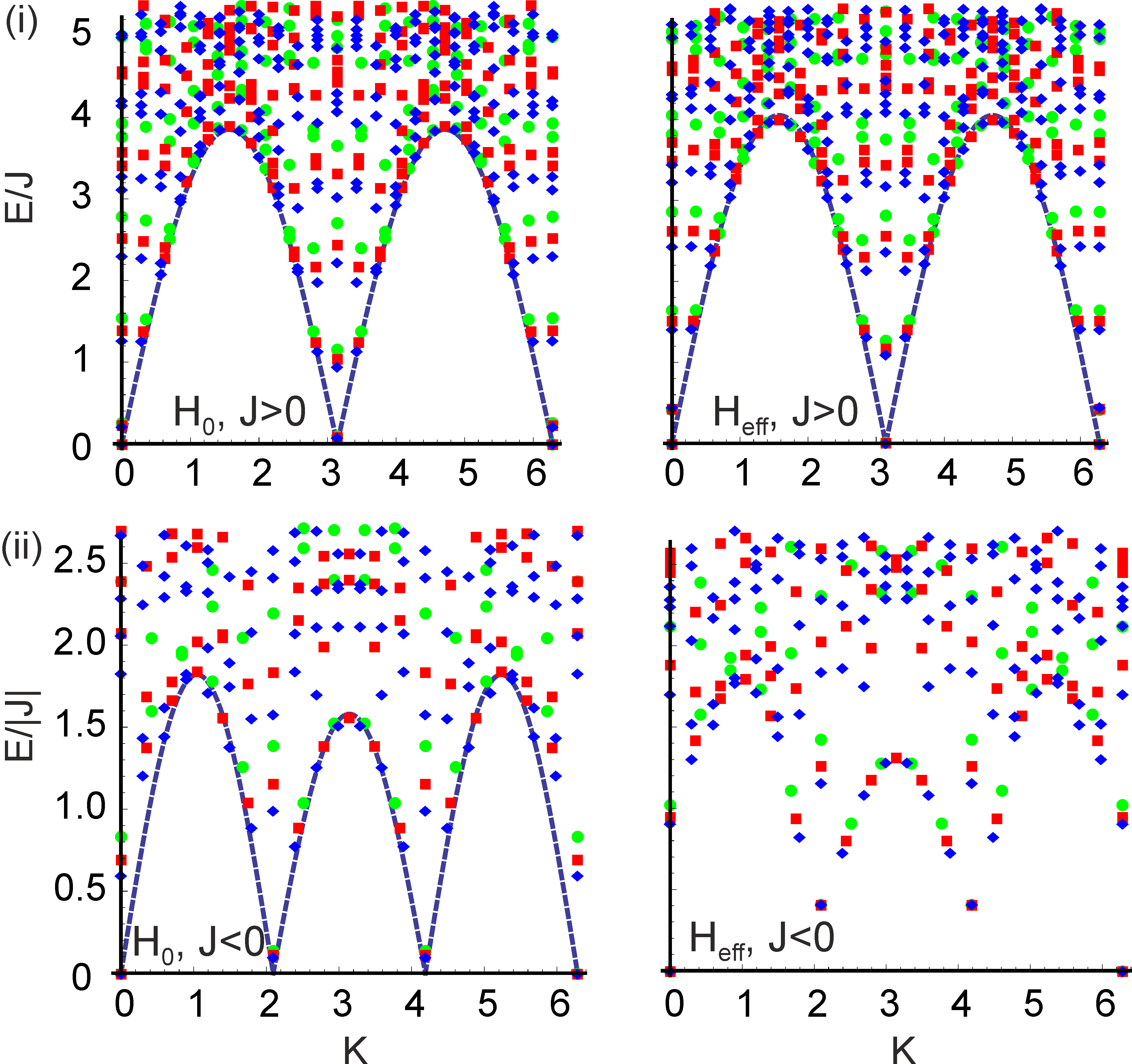}
\caption{\textbf{Low energy excitation spectra.} Energy spectra for $H_0$ (left column) and $H_{\rm eff}$ (right column) with positive and negative $J$ (referred to as case (i) and case (ii) in the text). The spectra have been shifted such that the lowest energy eigenvalues are zero. The dashed lines sketch low-energy dispersion relations and serve as a guide to the eye. The different symbols (green circles, red squares, blue diamonds) indicate different system sizes $N$: top panels (19,21,23), bottom panels (16,19,22).}
\label{fig:spectra}
\end{figure}
Let us finally analyze the ground state and the low-energy properties of the model. The ground state is in general a strongly correlated many-body quantum state whose preparation typically requires an adiabatic passage protocol: Here, one starts in a simple initial (product) state with no Rydberg atoms present and slowly tunes the laser parameters until one reaches the point (\ref{eq:parameters}) in parameter space. This has to be achieved in a time much smaller than the radiative lifetime of Rydberg atoms, which is typically (for principal quantum numbers in the range $n=40\ldots60$) on the order of $100\,\mu$s. The experimental feasibility of such protocol has recently been shown \cite{Schachenmayer10}. Beyond that it is moreover possible to infer properties of the ground state such as critical behavior from the time evolution and the scaling properties of certain quantum mechanical observables as demonstrated in Ref. \cite{Low09}.

The ground state phase diagram of the principal Hamiltonian $H_0$ has been studied in \cite{Fendley04}. It has a number of interesting features, e.g. for certain parameter choices the ground state becomes a simple matrix product state \cite{Lesanovsky11} and also the excitation gap can be calculated analytically \cite{Lesanovsky12}. There are furthermore integrable lines in parameter space that also include the point (\ref{eq:parameters}) which represents the anyonic Heisenberg interaction \cite{Feiguin07}. Here one can map the model onto the restricted-solid-on-solid model \cite{ABF84} and the critical properties of the ground state of $H_0$ are inferred from the exact solution. We consider the two cases (i) antiferromagnetic interaction between anyons ($J>0$, $V_2<0$)  and (ii) ferromagnetic interaction ($J<0$, $V_2>0$ ), separately. For case (i), the low-energy excitation is gapless and the continuum limit of the model is described by the minimal conformal field theory (CFT) with central charge $c=7/10$ \cite{Huse84}. For case (ii), the low-energy excitation is again gapless but the continuum limit is described by another CFT with $c=4/5$ \cite{Huse84}. The left column of Fig. \ref{fig:spectra} shows the low-energy spectra obtained by exact diagonalization up to $N=23$ with periodic boundary conditions. For (i), two gapless modes at momenta $K=0$ and $K=\pi$ are clearly visible. On the other hand, for (ii), the gapless modes are located at $K=0$, $2\pi/3$, and $4\pi/3$. These gapless modes suggest the existence of the quasi-long-range order. The difference of modulation period between (i) and (ii) can be understood as follows: In the original physical system, $J>0$ ($J<0$) and hence $V_2<0$ ($V_2>0$) means the attractive (repulsive) interaction between atoms in Rydberg states. But since there is a hard-core constraint caused by the Rydberg blockade, the density wave order of period $2$ is favored when $J>0$, while that of period $3$ is favored when $J<0$.

We finally discuss the effect of the hitherto neglected perturbations, i.e., $H_{\rm vdW}$ and $H_2$. Fig. \ref{fig:spectra} shows in the right column how the low-energy spectra are modified by these perturbations. The spectrum for case (i) with $H_{\rm vdW}$ and $H_2$ suggests that the gap is small and the model still remains in the vicinity of the critical point. As demonstrated in Ref. \cite{Feiguin07}, the gaplessness of the antiferromagnetic anyonic chain is protected by translation symmetry and topological ($Y$) symmetry. Since the first and the second terms in $H_2$ can be absorbed into $H_0$ by fine-tuning $\Delta$ and $V_2$, we expect that the criticality of the model is robust against the longer-range interactions stemming from the tail of the van-der-Waals potential and the correction due to the non-perfect blockade. On the other hand the spectrum for case (ii) clearly indicates the existence of a gap. The numerical results do not show any evidence of the quasi-degeneracy of the ground state caused by the density wave order. We thus conclude that the perturbations $H_{\rm vdW}$ and $H_2$ lead to a disordered gapped state that does not break any symmetries.

In conclusion, we have shown that a Rydberg lattice gas rather naturally constitutes an analog quantum simulator platform for Fibonacci anyons with tuneable interactions. Anyonic observables can be monitored by measuring atomic correlation functions which opens up the possibility for probing correlated quantum states of non-Abelian anyons in current experiments.

\acknowledgments
We acknowledge C. Ates, M. M\"uller, B. Olmos and S. Furukawa for fruitful discussions. H.K. was supported in part by Grant-in-Aid for Young Scientists (B) (23740298). I.L. acknowledges support by EPSRC and through the Leverhulme Trust.

\end{document}